\documentclass{article}
\usepackage{graphicx}
\usepackage{amsmath}

\newtheorem{theorem}{Theorem}

\input{tcilatex}

\begin{document}

\title{{\large Exact phase shifts for atom interferometry}}
\author{Ch. Antoine $^{1,a}$ and Ch.J. Bord\'{e} $^{2\text{,}a,b}$ \\
$^{a}${\small \ Equipe de Relativit\'{e} Gravitation et Astrophysique,
LERMA,CNRS-Observatoire de Paris,}\\
{\small Universit\'{e} Pierre et Marie Curie, 4 place Jussieu, 75005 Paris,
France}\\
{\small \ }$^{b}${\small \ Laboratoire de Physique des Lasers, UMR 7538
CNRS,\ Universit\'{e} Paris Nord,}\\
{\small 99 avenue J.-B. Cl\'{e}ment, 93430 Villetaneuse, France}}
\date{}
\maketitle

\begin{abstract}
In the case of an external Hamiltonian at most quadratic in position and
momentum operators, we use the ABCD$\xi $\ formulation of atom optics to
establish an exact analytical phase shift expression for atom
interferometers with arbitrary spatial or temporal beam splitter
configurations. This result is expressed in terms of coordinates and momenta
of the wave packet centers at the interaction vertices only.
\end{abstract}

\footnotetext[1]{
E-mail: antoinec@ccr.jussieu.fr} 
\footnotetext[2]{
E-mail: chbo@ccr.jussieu.fr}

\section{Introduction}

\bigskip

Recently atom interferometers \cite{berman} have been described by the ABCD$%
\xi $ formalism of Gaussian atom optics \cite{thtools}, \cite{metrologia}
which yields an exact formulation of phase shifts taking into account the
wave packet structure of atom waves.

For the theory of atom interferometers two basic stages are required:

1. a proper description of the propagation of wave packets between the beam
splitters

2. an adequate modelization of the beam splitters themselves.

The first stage is achieved through the ABCD$\xi $ theorem whose main
results are briefly recalled in section 2. The second problem is adressed by
the $ttt$ theorem which provides a simple model for the phase introduced by
the splitting process.

In this paper we give a compact way to express the atom interferometer phase
shifts in terms of the coordinates and momenta of the wave packet centers
only. For this purpose we derive two new theorems (the four end-points
theorem and the phase shift theorem) valid for a Hamiltonian
at-most-quadratic in position and momentum operators.

\bigskip

\section{The ABCD$\protect\xi $ theorem}

\bigskip

In this framework we consider a Hamiltonian which is the sum of an internal
Hamiltonian $H_{0}$ (with eigenvalues written with rest masses $m_{i}$) and
an external Hamiltonian $H_{ext}$:

\begin{equation}
H_{ext}=\frac{1}{2m}\overrightarrow{p_{op}}.\overset{\Rightarrow }{g}\left(
t\right) .\overrightarrow{p_{op}}-\frac{m}{2}\overrightarrow{q_{op}}.%
\overset{\Rightarrow }{\gamma }\left( t\right) .\overrightarrow{q_{op}}-%
\overrightarrow{\Omega }\left( t\right) .\left( \overrightarrow{q_{op}}%
\times \overrightarrow{p_{op}}\right) -m\overrightarrow{g}\left( t\right) .%
\overrightarrow{q_{op}}
\end{equation}
where one recognizes several usual gravito-inertial effects: rotation
through $\overrightarrow{\Omega }\left( t\right) $, gravity through $%
\overrightarrow{g}\left( t\right) $, gradient of gravity through \ $\overset{%
\Rightarrow }{\gamma }\left( t\right) $,... and where $\overset{\Rightarrow 
}{g}\left( t\right) $ is usually equal to the unity tensor (for simplicity
we omit the transposition sign $\sim $ on vectors).

For a wave packet $\psi \left( q,t_{1}\right) =wp\left(
q,t_{1},q_{1},p_{1},X_{1},Y_{1}\right) $, where $q_{1}$ is the initial mean
position of the wave-packet, $p_{1}$ its initial mean momentum, and $\left(
X_{1},Y_{1}\right) $ its initial complex width parameters in phase space,
one obtains the ABCD$\xi $ theorem \cite{thtools}: 
\begin{eqnarray}
\psi \left( q,t_{2}\right) &=&\int d^{3}q^{\prime }.K\left(
q,t_{2},q^{\prime },t_{1}\right) .wp\left( q^{\prime
},t_{1},q_{1},p_{1},X_{1},Y_{1}\right)  \notag \\
&=&e^{\frac{i}{\hbar }S_{cl}\left( t_{2},t_{1},q_{1},p_{1}\right) }.wp\left(
q,t_{2},q_{2},p_{2},X_{2},Y_{2}\right)
\end{eqnarray}
where $K$ and $S_{cl}$ are the quantum propagator and classical action
respectively, and where $q_{2},p_{2},X_{2},Y_{2}$ obey the ABCD\ law ($G$
and $U$ are the representative matrices of $\overset{\Rightarrow }{g}\left(
t\right) $ and the rotation $\overrightarrow{\Omega }\left( t\right) $, and
we write $A_{21}$ instead of $A\left( t_{2},t_{1}\right) $ for simplicity): 
\begin{equation}
\left( 
\begin{array}{c}
q_{2} \\ 
p_{2}/m
\end{array}
\right) =\left( 
\begin{array}{c}
U_{21}.\xi _{21} \\ 
G_{2}^{-1}.U_{21}.\dot{\xi}_{21}
\end{array}
\right) +\left( 
\begin{array}{cc}
A_{21} & B_{21} \\ 
C_{21} & D_{21}
\end{array}
\right) .\left( 
\begin{array}{c}
q_{1} \\ 
p_{1}/m
\end{array}
\right)  \label{q2p2}
\end{equation}
\begin{equation}
\left( 
\begin{array}{c}
X_{2} \\ 
Y_{2}
\end{array}
\right) =\left( 
\begin{array}{cc}
A_{21} & B_{21} \\ 
C_{21} & D_{21}
\end{array}
\right) .\left( 
\begin{array}{c}
X_{1} \\ 
Y_{1}
\end{array}
\right)
\end{equation}

For example the phase of a gaussian wave packet is: 
\begin{equation}
S_{cl}\left( t_{2},t_{1},q_{1},p_{1}\right) /\hbar +p_{2}.\left(
q-q_{2}\right) /\hbar +\frac{m}{2\hbar }\left( q-q_{2}\right) .\func{Re}%
\left( Y_{2}.X_{2}^{-1}\right) .\left( q-q_{2}\right)
\end{equation}
\bigskip and in this case the main phase shift recorded between $t_{1}$ and $%
t_{2}$ is equal to: 
\begin{equation}
S_{cl}\left( t_{2},t_{1},q_{1},p_{1}\right) /\hbar +p_{1}.q_{1}/\hbar
-p_{2}.q_{2}/\hbar
\end{equation}

\bigskip

\section{The ttt theorem}

\bigskip

When the dispersive nature of a laser beam splitter is neglected (wave
packets structure preserved), its effect may be summarized by the
introduction of both a phase and an amplitude factor (see \cite{ishikawa}
and \cite{ttt} for a detailed proof):

\begin{equation}
M_{ba}.e^{-i\left( \omega t^{\ast }-k^{\ast }.q^{\ast }+\varphi ^{\ast
}\right) }
\end{equation}
where $t^{\ast }$ and $q^{\ast }$ depend on $t_{A}$ and $q_{A}$, the mean
time and position of the laser wave.

For a temporal beam splitter: 
\begin{eqnarray}
t^{\ast } &\equiv &t_{A}  \notag \\
q^{\ast } &\equiv &q_{cl}\left( t_{A}\right)  \notag \\
k^{\ast } &\equiv &k  \notag \\
\varphi ^{\ast } &\equiv &\varphi \text{ (laser phase)}
\end{eqnarray}

For a spatial beam splitter: 
\begin{eqnarray}
q^{\ast } &\equiv &q_{A}  \notag \\
t^{\ast }\text{ such that }q_{cl}\left( t^{\ast }\right) &\equiv &q_{A} 
\notag \\
k^{\ast } &\equiv &k+\delta k  \notag \\
\varphi ^{\ast } &\equiv &\varphi +\delta \varphi
\end{eqnarray}
where $\delta k$ is the additional momentum transferred to the excited atoms
out of resonance, and where $\delta \varphi $ is a laser phase: $\delta
\varphi \equiv -\delta k.q_{A}$ (see \cite{ttt}).

Let us emphasize that these calculations do not rely on the assumption that
the splitter is infinitely thin or that the atom trajectories are classical.

\bigskip

\section{The four end-points theorem for a Hamiltonian at most quadratic in
position and momentum operators}

\bigskip

We shall cut any interferometer into as many slices as there are
interactions on either arm and thus obtain several path pieces (see section
5). From now on we shall consider systematically pairs of these homologous
paths (see Fig. 1) in the case of a Hamiltonian at most quadratic.

\FRAME{dtbpFU}{208.75pt}{115.3125pt}{0pt}{\Qcb{Fig. 1: A pair of homologous
paths}}{\Qlb{fig1}}{Figure }{\special{language "Scientific Word";type
"GRAPHIC";display "USEDEF";valid_file "T";width 208.75pt;height
115.3125pt;depth 0pt;original-width 397.5pt;original-height
219.0625pt;cropleft "0";croptop "1";cropright "1";cropbottom
"0";tempfilename 'H309HI07.wmf';tempfile-properties "XPR";}}

These two classical trajectories are labelled by their corresponding mass ($%
m_{\alpha }$ and $m_{\beta }$), their initial position and momentum ($%
q_{\alpha 1}$, $p_{\alpha 1}$, $q_{\beta 1}$ and $p_{\beta 1}$) and their
common drift time $T=t_{2}-t_{1}$.

Before establishing the first new theorem let us consider the expression of
the classical action for the $\alpha $ path (see \cite{thtools}): 
\begin{eqnarray}
\frac{S_{cl}\left( t_{2},t_{1},q_{\alpha 1},p_{\alpha 1}\right) }{m_{\alpha }%
} &=&\dot{\xi}.\widetilde{U}.G^{-1}.\left( A.q_{\alpha 1}+B.p_{\alpha
1}/m_{\alpha }\right) +\int_{t_{1}}^{t_{2}}\frac{L}{m_{\alpha }}dt  \notag \\
&&+q_{\alpha 1}.\frac{\widetilde{A}C}{2}.q_{\alpha 1}+\frac{p_{\alpha 1}}{%
m_{\alpha }}.\frac{\widetilde{B}D}{2}.\frac{p_{\alpha 1}}{m_{\alpha }}+\frac{%
p_{\alpha 1}}{m_{\alpha }}.\widetilde{B}C.q_{\alpha 1}
\end{eqnarray}
which can be rewritten as: 
\begin{eqnarray}
\frac{S_{cl}\left( t_{2},t_{1},q_{\alpha 1},p_{\alpha 1}\right) }{m_{\alpha }%
} &=&\frac{p_{\alpha 2}}{2m_{\alpha }}.q_{\alpha 2}-\frac{p_{\alpha 1}}{%
2m_{\alpha }}.q_{\alpha 1}-\frac{1}{2}\dot{\xi}.\widetilde{U}.G^{-1}.U.\xi
+\int_{t_{1}}^{t_{2}}\frac{L}{m_{\alpha }}dt  \notag \\
&&+\frac{1}{2}\dot{\xi}.\widetilde{U}.G^{-1}.q_{\alpha 2}-\frac{1}{2}\xi .%
\widetilde{U}.\frac{p_{\alpha 2}}{m_{\alpha }}
\end{eqnarray}
with the help of the definition of $q_{\alpha 2}$ and $\frac{p_{\alpha 2}}{%
m_{\alpha }}$ (see \ref{q2p2}). Then we can use the $\beta $ path to replace 
$\dot{\xi}.\widetilde{U}.G^{-1}$\ with: 
\begin{equation}
\dot{\xi}.\widetilde{U}.G^{-1}=\frac{p_{\beta 2}}{m_{\beta }}-C.q_{\beta
1}-D.\frac{p_{\beta 1}}{m_{\beta }}
\end{equation}

Consequently we get: 
\begin{eqnarray}
\frac{S_{cl}\left( t_{2},t_{1},q_{\alpha 1},p_{\alpha 1}\right) }{m_{\alpha }%
} &=&\frac{1}{2}\left( \frac{p_{\alpha 2}}{m_{\alpha }}+\frac{p_{\beta 2}}{%
m_{\beta }}\right) .q_{\alpha 2}-\frac{1}{2}\left( \frac{p_{\alpha 1}}{%
m_{\alpha }}+\frac{p_{\beta 1}}{m_{\beta }}\right) .q_{\alpha 1}  \notag \\
&&+h\left( t_{2},t_{1}\right) +f\left( \alpha ,\beta \right)
\end{eqnarray}
where $h\left( t_{2},t_{1}\right) $ is independent of positions and momenta
and where $f\left( \alpha ,\beta \right) =f\left( \beta ,\alpha \right) $.
The same goes for the expression of $S_{cl}\left( t_{2},t_{1},q_{\beta
1},p_{\beta 1}\right) /m_{\beta }$ which is obtained by exchanging $\alpha $
and $\beta $. Finally we arrive at the first new theorem (a more general
demonstration starting with Hamilton principal functions is given in
appendix 1):

\begin{theorem}
\begin{eqnarray}
&&\frac{S_{cl}\left( t_{2},t_{1},q_{\alpha 1},p_{\alpha 1}\right) }{%
m_{\alpha }}-\frac{1}{2}\left( \frac{p_{\alpha 2}}{m_{\alpha }}+\frac{%
p_{\beta 2}}{m_{\beta }}\right) .q_{\alpha 2}+\frac{1}{2}\left( \frac{%
p_{\alpha 1}}{m_{\alpha }}+\frac{p_{\beta 1}}{m_{\beta }}\right) .q_{\alpha
1}  \notag \\
&=&\frac{S_{cl}\left( t_{2},t_{1},q_{\beta 1},p_{\beta 1}\right) }{m_{\beta }%
}-\frac{1}{2}\left( \frac{p_{\alpha 2}}{m_{\alpha }}+\frac{p_{\beta 2}}{%
m_{\beta }}\right) .q_{\beta 2}+\frac{1}{2}\left( \frac{p_{\alpha 1}}{%
m_{\alpha }}+\frac{p_{\beta 1}}{m_{\beta }}\right) .q_{\beta 1}
\end{eqnarray}
\end{theorem}

or equivalently: 
\begin{eqnarray}
&&\frac{S_{cl}\left( t_{2},t_{1},q_{\alpha 1},p_{\alpha 1}\right) }{%
m_{\alpha }}-\frac{p_{\alpha 2}}{m_{\alpha }}.q_{\alpha 2}+\frac{p_{\alpha 1}%
}{m_{\alpha }}.q_{\alpha 1}  \notag \\
&&-\left[ \frac{S_{cl}\left( t_{2},t_{1},q_{\beta 1},p_{\beta 1}\right) }{%
m_{\beta }}-\frac{p_{\beta 2}}{m_{\beta }}.q_{\beta 2}+\frac{p_{\beta 1}}{%
m_{\beta }}.q_{\beta 1}\right]  \notag \\
&=&\left( \frac{p_{\beta 2}}{m_{\beta }}-\frac{p_{\alpha 2}}{m_{\alpha }}%
\right) .\left( \frac{q_{\alpha 2}+q_{\beta 2}}{2}\right) -\left( \frac{%
p_{\beta 1}}{m_{\beta }}-\frac{p_{\alpha 1}}{m_{\alpha }}\right) .\left( 
\frac{q_{\alpha 1}+q_{\beta 1}}{2}\right)
\end{eqnarray}
which gives the main part of the phase shift expressed with the half sums of
the coordinates and the momenta of the four end-points only.

In the case of identical masses ($m_{\alpha }=m_{\beta }$) this expression
simplifies to: 
\begin{eqnarray}
&&S_{cl}\left( t_{2},t_{1},q_{\alpha 1},p_{\alpha 1}\right) -p_{\alpha
2}.q_{\alpha 2}+p_{\alpha 1}.q_{\alpha 1}  \notag \\
&&-\left[ S_{cl}\left( t_{2},t_{1},q_{\beta 1},p_{\beta 1}\right) -p_{\beta
2}.q_{\beta 2}+p_{\beta 1}.q_{\beta 1}\right]  \notag \\
&=&\left( p_{\beta 2}-p_{\alpha 2}\right) .\left( \frac{q_{\alpha
2}+q_{\beta 2}}{2}\right) -\left( p_{\beta 1}-p_{\alpha 1}\right) .\left( 
\frac{q_{\alpha 1}+q_{\beta 1}}{2}\right)
\end{eqnarray}

\bigskip

\section{The phase shift theorem for a Hamiltonian at most quadratic in
position and momentum operators}

\bigskip

In this section we draw on the results of previous sections to establish the
interferometer phase shift expression for an arbitrary beam splitters
configuration.

For a sequence of pairs of homologous paths (an interferometer geometry)
(see Fig. 2)\FRAME{dtbpFU}{272.6875pt}{175pt}{0pt}{\Qcb{Fig. 2:
Interferometer geometry sliced into pairs of homologous paths between
interactions on either arm (when an interaction occurs only on one arm the
corresponding \textit{k} on the other arm is set = 0)}}{\Qlb{fig2}}{Figure }{%
\special{language "Scientific Word";type "GRAPHIC";display
"USEDEF";valid_file "T";width 272.6875pt;height 175pt;depth
0pt;original-width 474.375pt;original-height 363.875pt;cropleft "0";croptop
"1";cropright "1";cropbottom "0";tempfilename
'H309X708.wmf';tempfile-properties "XPR";}}one can infer the general sum for
the main coordinate dependant part of the global phase shift:

\begin{equation}
\frac{p_{\beta D}-p_{\alpha D}}{\hbar }.\left( q-\frac{q_{\alpha D}+q_{\beta
D}}{2}\right) -\frac{p_{\alpha 1}+p_{\beta 1}}{2\hbar }.\left( q_{\beta
1}-q_{\alpha 1}\right) +\sum_{i=1}^{N}\left( k_{\beta i}-k_{\alpha i}\right)
.\frac{q_{\alpha i}+q_{\beta i}}{2}
\end{equation}

If now we take into account the other terms of the phase shift we finally
obtain the following result (given here for a Gaussian wave packet):

\begin{theorem}
\begin{eqnarray}
\Delta \phi \left( q,t_{N+1}\equiv t_{D}\right)  &=&\left( p_{\beta
D}-p_{\alpha D}\right) .\left( q-\frac{q_{\alpha D}+q_{\beta D}}{2}\right)
/\hbar -\frac{p_{\alpha 1}+p_{\beta 1}}{2\hbar }.\left( q_{\beta
1}-q_{\alpha 1}\right)   \notag \\
&&+\sum_{i=1}^{N}\left[ \left( k_{\beta i}-k_{\alpha i}\right) .\frac{%
q_{\alpha i}+q_{\beta i}}{2}-\left( \omega _{\beta i}-\omega _{\alpha
i}\right) .t_{i}+\varphi _{\beta i}-\varphi _{\alpha i}\right]   \notag \\
&&+\sum_{i=1}^{N}\left( \frac{m_{\beta i}}{m_{\alpha i}}-1\right) .  \notag
\\
&&\left[ \frac{S_{\alpha i}}{\hbar }+\frac{p_{\alpha ,i+1}}{2\hbar }.\left(
q_{\beta ,i+1}-q_{\alpha ,i+1}\right) -\frac{p_{\alpha i}+\hbar k_{\alpha i}%
}{2\hbar }.\left( q_{\beta i}-q_{\alpha i}\right) \right]   \notag \\
&&+\frac{m_{\beta ,N}}{2\hbar }\left( q-q_{\beta D}\right) .\func{Re}\left(
Y_{D}.X_{D}^{-1}\right) .\left( q-q_{\beta D}\right)   \notag \\
&&-\frac{m_{\alpha ,N}}{2\hbar }\left( q-q_{\alpha D}\right) .\func{Re}%
\left( Y_{D}.X_{D}^{-1}\right) .\left( q-q_{\alpha D}\right) 
\end{eqnarray}
\end{theorem}

where $S_{\alpha i}\equiv S_{cl}\left( t_{i+1},t_{i},q_{\alpha i},p_{\alpha
i}+\hbar k_{\alpha i},m_{\alpha i}\right) $.

This fundamental result is valid for a time-dependent Hamiltonian and takes
into account all the mass differences which may occur. It allows to
calculate exactly the phase shift for all the interferometer geometries
which can be sliced as above: symmetrical Ramsey-Bord\'{e} (Mach-Zehnder),
atomic fountain clocks,... All these particular cases will be detailed in a
forthcoming paper (see \cite{lunteren})

Let us point out that the nature (temporal or spatial) of beam splitters
leads to different slicing of the paths. In the spatial case, indeed, the
number of different $t_{i}^{\ast }$ may be twice as great as in the temporal
case (see the definition of $t_{i}^{\ast }$ in these two different cases in
section 3).

\bigskip

\section{Phase shift after spatial integration}

\bigskip

In an actual interferometer one has to integrate spatially the output wave
packet over the detection region. With Gaussian wave packets this
integration leads to a mid-point theorem \cite{metrologia} \cite{lunteren}:
''\textit{The first term of }$\Delta \phi \left( q,t_{D}\right) $\textit{\
disappears when the spatial integration is performed''.}

Furthermore the terms which depend on the wave packets structure ($Y$ and $X$%
) vanish when $m_{\beta ,N}=m_{\alpha ,N}$ (which is always the case). One
obtains finally:

\begin{eqnarray}
\Delta \phi \left( t_{D}\right)  &=&-\frac{p_{\alpha 1}+p_{\beta 1}}{2\hbar }%
.\left( q_{\beta 1}-q_{\alpha 1}\right)   \notag \\
&&+\sum_{i=1}^{N}\left[ \left( k_{\beta i}-k_{\alpha i}\right) .\frac{%
q_{\alpha i}+q_{\beta i}}{2}-\left( \omega _{\beta i}-\omega _{\alpha
i}\right) .t_{i}+\varphi _{\beta i}-\varphi _{\alpha i}\right]   \notag \\
&&+\sum_{i=1}^{N}\left( \frac{m_{\beta i}-m_{\alpha i}}{2\hbar }\right) . \\
&&\{\left( \frac{S_{\alpha i}}{m_{\alpha i}}+\frac{p_{\alpha ,i+1}}{%
2m_{\alpha i}}.\left( q_{\beta ,i+1}-q_{\alpha ,i+1}\right) -\frac{p_{\alpha
i}+\hbar k_{\alpha i}}{2m_{\alpha i}}.\left( q_{\beta i}-q_{\alpha i}\right)
\right)   \notag \\
&&+\left( \frac{S_{\beta i}}{m_{\beta i}}+\frac{p_{\beta ,i+1}}{2m_{\beta i}}%
.\left( q_{\alpha ,i+1}-q_{\beta ,i+1}\right) -\frac{p_{\beta i}+\hbar
k_{\beta i}}{2m_{\beta i}}.\left( q_{\alpha i}-q_{\beta i}\right) \right) \}
\notag
\end{eqnarray}

\bigskip

\section{Identical masses and symmetrical case}

\bigskip

The case of identical masses is an important approximation which is commonly
used for the modelization of many devices like gravimeters and gyrometers 
\cite{chu2001}, \cite{kasevitch2000}, \cite{wolftourrenc}. If $m_{\alpha
i}=m_{\beta i}=m,$ $\forall i$, this general phase shift becomes : 
\begin{eqnarray}
\Delta \phi \left( t_{D}\right)  &=&-\frac{p_{\alpha 1}+p_{\beta 1}}{2\hbar }%
.\left( q_{\beta 1}-q_{\alpha 1}\right) +\sum_{i=1}^{N}\left( k_{\beta
i}-k_{\alpha i}\right) .\frac{q_{\alpha i}+q_{\beta i}}{2}  \notag \\
&&+\sum_{i=1}^{N}\left[ \varphi _{\beta i}-\varphi _{\alpha i}-\left( \omega
_{\beta i}-\omega _{\alpha i}\right) .t_{i}\right] 
\end{eqnarray}

We can also specify the form of this phase shift when the interferometer
geometry is symmetrical (see Fig.3).

\FRAME{dtbpFU}{301.875pt}{173.1875pt}{0pt}{\Qcb{Fig. 3: A typical
symmetrical interferometer}}{\Qlb{fig3}}{Figure }{\special{language
"Scientific Word";type "GRAPHIC";display "USEDEF";valid_file "T";width
301.875pt;height 173.1875pt;depth 0pt;original-width
580.9375pt;original-height 291.3125pt;cropleft "0";croptop "1";cropright
"1.0016";cropbottom "0";tempfilename 'H308XI05.wmf';tempfile-properties
"XPR";}}

This symmetry is expressed as : $k_{\beta i}+k_{\alpha i}=0,$ $\forall i\in %
\left[ 2,N-1\right] $, i.e. it is a symmetry with respect to the direction
of the particular vector : $p_{initial}+\hbar k_{initial}/2$.

Consequently : 
\begin{equation}
\Delta \phi \left( t_{N}\right) =k_{1}.q_{1}+2\sum_{i=2}^{N-1}k_{i}.\frac{%
q_{\alpha i}+q_{\beta i}}{2}+k_{N}.\frac{q_{\alpha N}+q_{\beta N}}{2}%
-\sum_{i=1}^{N-1}\left( \varphi _{\beta i}-\varphi _{\alpha i}\right)
\end{equation}

But $\forall i\in \left[ 2,N-1\right] $ : 
\begin{eqnarray}
\text{ }\frac{q_{\alpha ,i+1}+q_{\beta ,i+1}}{2} &=&\xi _{i+1,i}+A_{i+1,i}.%
\frac{q_{\alpha i}+q_{\beta i}}{2}+\frac{B_{i+1,i}}{m}.\frac{p_{\alpha
i}+p_{\beta i}}{2}  \notag \\
&=&\xi _{i+1,1}+A_{i+1,1}.q_{1}+\frac{B_{i+1,1}}{m}.(p_{1}+\frac{\hbar k_{1}%
}{2})  \notag \\
&\equiv &Q\left( t_{i+1}\right)
\end{eqnarray}
which can be calculated with the ABCD$\xi $ law.

It depends only on $q_{1}$ (initial position) and $p_{1}+\frac{\hbar k_{1}}{2%
}$ (``Bragg initial momentum'').

Therefore : 
\begin{equation}
\Delta \phi \left( t_{N}\right) =\sum_{i=1}^{N}(k_{\beta i}-k_{\alpha
i}).Q\left( t_{i}\right) -\sum_{i=1}^{N-1}\left( \varphi _{\beta i}-\varphi
_{\alpha i}\right)
\end{equation}
which has a very simple form when the origin of coordinates is chosen such
that $q_{1}=0$, and when the Bragg condition $p_{1}+\frac{\hbar k_{1}}{2}=0$
is satisfied.

\bigskip

\section{Conclusion}

\bigskip

In this paper we have used the ABCD$\xi $\ formulation of atom optics and
the $ttt$ theorem to establish two theorems valid for a time-dependent
Hamiltonian at most quadratic in position and momentum operators. The first
one gives a compact expression of the action difference between two
homologous paths. The second one gives an analytical expression of the
global phase shift for atom interferometers in the case of such a
Hamiltonian.

Consequently this analytical expression provides a simple way to calculate
exactly the phase shift in this case, and then to calculate perturbatively
for example the effect of a third-order term in the external Hamiltonian
(necessary for space missions like HYPER \cite{hyper}). For example, one can
calculate exactly the global phase shift due to gravity plus a gradient of
gravity plus a rotation, and then calculate perturbatively the effect of a
gradient of gradient of gravity. These calculations and the application to
specific cases (gravimeters, gyrometers, atomic clocks...) will be detailed
in a forthcoming article \cite{lunteren} where we recover well-known
perturbative results (\cite{audretsch}, \cite{wolftourrenc}, \cite{pla89}, 
\cite{spectro10}) from exact expressions.

\bigskip

\section{Appendix 1}

\bigskip

In a case of a Hamiltonian at most quadratic in position and momentum
operators, the Hamilton principal functions concerning two pairs of
homologous points are also at most quadratic in positions (owing to the
Hamilton-Jacobi equation, see \cite{thtools}): 
\begin{equation}
S_{\alpha }\left( q_{\alpha 1},q_{\alpha 2}\right) /m_{\alpha
}=a+b.q_{\alpha 1}+c.q_{\alpha 2}+q_{\alpha 1}.d.q_{\alpha 1}+q_{\alpha
1}.e.q_{\alpha 2}+q_{\alpha 2}.f.q_{\alpha 2}
\end{equation}
\begin{equation}
S_{\beta }\left( q_{\beta 1},q_{\beta 2}\right) /m_{\beta }=a+b.q_{\beta
1}+c.q_{\beta 2}+q_{\beta 1}.d.q_{\beta 1}+q_{\beta 1}.e.q_{\beta
2}+q_{\beta 2}.f.q_{\beta 2}
\end{equation}
where $a$ is a scalar, $b$ and $c$ are vectors, and $d$, $e$ and $f$ are
matrices (see \cite{thtools}).

We can define $p_{\alpha 1}$, $p_{\alpha 2}$, $p_{\beta 1}$, $p_{\beta 2}$
such that: 
\begin{equation}
p_{\alpha 1}\equiv -\nabla _{q_{\alpha 1}}\left( \frac{S_{\alpha }}{%
m_{\alpha }}\right) =-b-2d.q_{\alpha 1}-e.q_{\alpha 2}
\end{equation}
\begin{equation}
p_{\alpha 2}\equiv \nabla _{q_{\alpha 2}}\left( \frac{S_{\alpha }}{m_{\alpha
}}\right) =c+2f.q_{\alpha 2}+\widetilde{e}.q_{\alpha 1}
\end{equation}
\begin{equation}
p_{\beta 1}\equiv -\nabla _{q_{\beta 1}}\left( \frac{S_{\beta }}{m_{\beta }}%
\right) =-b-2d.q_{\beta 1}-e.q_{\beta 2}
\end{equation}
\begin{equation}
p_{\beta 2}\equiv \nabla _{q_{\beta 2}}\left( \frac{S_{\beta }}{m_{\beta }}%
\right) =c+2f.q_{\beta 2}+\widetilde{e}.q_{\beta 1}
\end{equation}
and obtain the following expression: 
\begin{equation}
\frac{S_{\alpha }}{m_{\alpha }}-\frac{S_{\beta }}{m_{\beta }}=\frac{1}{2}%
\left( p_{\alpha 2}+p_{\beta 2}\right) .\left( q_{\alpha 2}-q_{\beta
2}\right) -\frac{1}{2}\left( p_{\alpha 1}+p_{\beta 1}\right) .\left(
q_{\alpha 1}-q_{\beta 1}\right) 
\end{equation}

The same relation holds for the classical action concerning two actual paths
with a common drift time (homologous paths). This yields an other
demonstration of the first theorem expressed in section 4.

\end{document}